\documentclass[prl, twocolumn, showpacs]{revtex4}
\usepackage{amssymb}
\usepackage{amsmath}
\usepackage{epsfig}
\usepackage{bm}

\begin{document}

\title{ Two-dimensional  electron honey: highly viscous electron fluid \\
in which   transverse magnetosonic waves can propagate }
\author{ P. S. Alekseev and A. P. Alekseeva }
\affiliation{Ioffe  Institute,  194021  St.~Petersburg, Russia}

\begin{abstract}

One of the main macroscopic differences between ordinary and highly viscous fluids
 is the lack of transverse sound in the first and possibility of its excitation
 in the second. In modern high-mobility conductors (Weyl semimetals,
  best-quality quantum wells, and graphene) electrons
 can form a viscous fluid at low temperatures.
In this work we develop high-frequency hydrodynamics of
 two-dimensional highly viscous electron  fluids in magnetic field.
Such fluids are characterized by simultaneous presence
 of the  excitations associated with the elastic stress (transverse sound)
  as well as with the violation of the local charge neutrality (plasmons).
  We demonstrate that  both the viscoelastic and the plasmonic
  components of a flow can  exhibit
   the {\em viscous resonance}   that was recently proposed
  for charged viscous fluids.  This resonance   is related to rotation
 of the viscous stress tensor of a charged  fluid in magnetic field.
   We argue that the viscous resonance
   is responsible for the peak and the peculiarities
   observed in photoresistance and photovoltage of the
   ultra-high mobility  GaAs quantum wells.  We conclude
   that a highly viscous electron fluid
   (``electron honey'')  is realized in those structures.


\end{abstract}

\maketitle

{\em 1. Introduction.}
In materials with enough weak disorder
 phonons and conduction electrons can form
 viscous  fluids provided that
  the inter-particle  collisions conserving  momentum
 are much more intensive than any other collisions
 which do not conserve  momentum \cite{Gurzhi}.
 The hydrodynamic regime
 of phonon transport in liquid helium and dielectrics
    was studied several decades ago
    in sufficient details \cite{Gurevich,Pitajevkii}.
 However, only recently the hydrodynamic regime
  of charge transport was discovered
  in the novel ultra-high quality materials: GaAs quantum wells
    \cite{exps_neg_1,exps_neg_2,exps_neg_3,exps_neg_4,Gusev_1,Gusev_2},
    2D monovalent layered metal PdCoO$_2$ \cite{Weyl_sem_1},
3D Weyl semimetal WP$_2$ \cite{Weyl_sem_2},
    and   graphene \cite{grahene,grahene_2,grahene_3}.
  These experimental discoveries were accompanied by
  an extensive development of theory
   \cite{Spivak,Andreev,Mendoza_Herrmann_Succi,
   Tomadin_Vignale_Polini,je_visc,Levitov_et_al,Levitov_et_al_2,
   Lucas,Lucas_2,eta_xy,we_4,we_5_1,we_5_2,we_5,recentest_,vis_res,recentest,
   recentest2,recentest3,L_n_1,L_n_2,Khoo_Villadiego}.

    One of the evidences   of the hydrodynamic regime of electron  transport
     is the giant negative magnetoresistance  effect,
   which  is the decrease of resistance by 1-2
orders of magnitude in moderate magnetic fields,  was  discovered in
   the best-quality GaAs quantum wells
   \cite{exps_neg_1,exps_neg_2,exps_neg_3,exps_neg_4},
    the  Weyl semimetal  WP$_2$ \cite{Weyl_sem_2},
     and, very recently, in graphene \cite{grahene_3}.
This effect was explained  by the hydrodynamic model
  taking into account the dependence of  the viscosity
  coefficients of electrons on magnetic field and temperature \cite{je_visc}.
Another evidence of forming a viscous electron fluid in
graphene and high-mobility GaAs quantum wells is
 observation of  the negative nonlocal resistance
 related to formation of whirlpools
   \cite{grahene,Gusev_2,Levitov_et_al}.

 High-frequency transport in a viscous electron fluid
  was theoretically considered
 in several recent publications \cite{recentest,recentest2,recentest3,vis_res}.
  An ac  flow of the fluid in a long sample in zero magnetic field
  was studied in Refs.~\cite{recentest,recentest2}.
  A Navier-Stocks equation
 for the  two-dimension (2D) electron fluid in magnetic field
 with taking into account
  the time dispersion of viscosity was derived  in Refs.~\cite{eta_xy,vis_res}.
 The ac viscosity  coefficients of 2D electrons have a resonance
 at a frequency  $\omega$ equal to the doubled electron
 cyclotron frequency $2\omega_c$~\cite{vis_res}.
 Such the {\em viscous resonance } manifests itself
  in the damping coefficient of magnetoplasmons \cite{vis_res}.
Plasmons and transverse zero sound in a 2D electron Fermi liquid
 in zero magnetic field were studied in Refs.~\cite{L_n_1,Khoo_Villadiego}

 In this work we demonstrate that the viscous resonance
 can be used to detect the excitation of  transverse zero sound
   in a highly viscous electron fluid.
  We provide the evidences that this phenomenon have been observed
 in recent experiments
 on the 2D electron fluid in
 the best-quality GaAs quantum wells
 \cite{exp_GaAs_ac_1,exp_GaAs_ac_2,exp_GaAs_ac_3}.

An ac flow of a highly viscous electron fluid in a microwave
electric field consists  of the plasmonic part formed
  by standing waves of plasmons and the viscoelastic  part formed
  by standing waves of transverse zero sound.
    The last  exists in Fermi liquids with
    enough strong interaction  between quasi-particles \cite{LP}.
  We study transverse zero sound in a 2D electron Fermi liquid
   in magnetic field. We use the hydrodynamic approximation, which
   is valid at strong enough  inter-particle interaction.
    The viscoelastic contribution  determine the flow
 in the near-edge regions of wide samples
  and in the whole space in narrow samples~\cite{recentest,recentest2}.
  We demonstrate that in a perpendicular magnetic field the standing waves of
  the transverse zero sound    are formed    at the frequencies
    above the viscous resonance frequency, $\omega>2\omega_c$,
    and are not formed at $\omega<2\omega_c$.
    Because of this change of the flow,
     an ac current in narrow samples exhibits
  the viscous resonance at $\omega=2\omega_c$.
 Around these frequencies the behavior of the  electron fluid
 resembles   one of an amorphous elastic media
 near the viscoelasticity transition, so the term ``electron honey''
  can be coined.

We discuss the giant peak and the peculiarities that were
recently observed in photoresistance and  photovoltage
 of ultra-high mobility GaAs quantum wells at the frequencies
near $\omega  = 2 \omega_c$ \cite{exp_GaAs_ac_1,exp_GaAs_ac_2,exp_GaAs_ac_3}.
  We argue that  excitation of transverse zero sound,
 leading to the viscous resonance, is  responsible
  the observed phenomena. We also discuss that  the independence
  of the viscoelastic part of the flow
  on the sigh of the circular polarization
  of radiation correlates with the experiments
  \cite{Smet_1,Smet_2,Ganichev_1,Ganichev_2}.
  We believe that the phenomena studied  in this work should be expected
  in  high-quality graphene and other material
 where the viscous transport have been recently realized.

{\em 2. Ac hydrodynamics of 2D electron  Fermi gas
and strongly non-ideal Fermi liquid.}
If the  inter-particle interaction of electrons is weak and
they can be regarded as an almost ideal Fermi gas,
the hydrodynamic approach can be used
when the characteristic space scale
$W$ of changing of the hydrodynamic velocity $\mathbf{V}(\mathbf{r},t)$
 is far greater than, at least, one of the following lengths:
the electron mean  free path relative to electron-electron  collisions
$l_{ee}=v_F \tau_{ee}$; the electron cyclotron radius $R_c=v_F/\omega_c$;
the length of the path that free electron passes during
the period of changing of $\mathbf{V}(\mathbf{r},t)$,
$l_{\omega}=v_F/\omega$ \cite{Gurzhi}. Here  $\mathbf{r}=(x,y)$
is the coordinate in the 2D layer, $v_F$ is the Fermi velocity,
 $\omega$ is a characteristic frequency of a flow, $\omega_c$
 is the cyclotron frequency, and $\tau_{ee} $
 is the electron-electron scattering time.

 An ac flow of the electron gas is described by the particle
 density $n(\mathbf{r},t) = n_0 + \delta n(\mathbf{r},t) $
 and the hydrodynamic velocity $\mathbf{V}(\mathbf{r},t)$
 ($n_0$ is the equilibrium density).
We decompose  $\delta n(\mathbf{r},t)$ and  $\mathbf{V}(\mathbf{r},t)$
 by time harmonics proportional to $e^{-i\omega t }$
with the complex amplitudes $ \delta n(\mathbf{r} )  $ and $\mathbf{V} (\mathbf{r} ) $.
In the regime linear by the external ac electric field,
 the continuity equation and the Navier-Stocks equation
 in magnetic field $\mathbf{B} =B \mathbf{e}_z$
at the frequencies $\omega$  compared with $\omega_c$ and $1/\tau_{ee}$
have the form \cite{vis_res,SI}:
\begin{equation}
\label{cont_0}
- i\omega \, \delta  n + n_0\mathrm{div }  \mathbf{V} = 0
 \:,
 \end{equation}
 \begin{equation}
\label{Navier_Stocks_B_with_time_disp_0}
-i\omega  \mathbf{V}
=
 e\mathbf{E} /m
  + \omega_c \mathbf{V} \times \mathbf{e}_z
+ \eta_{xx} \, \Delta \mathbf{V}+  \eta _{xy} \,
   \Delta \mathbf{V} \times \mathbf{e}_z
\:,
\end{equation}
where $\mathbf{E} = \mathbf{E}(\mathbf{r})$
 is the complex amplitude of the electric field $\mathbf{E} (\mathbf{r},t)$,
 $e$ is the electron charge,  $m$ is the electron mass,
and the ac viscosity coefficients
$ \eta_{xx} = \eta_{xx} (\omega)$ and
 $ \eta_{xy} = \eta_{xy} (\omega)$ have the form \cite{vis_res}:
\begin{equation}
\label{eta zeta ot omega_0}
\begin{array}{c}
\eta_{xx}
\\
\eta_{xy}
\end{array}
\Big\}
=
\frac{ \eta_0 }
 {1+(-\omega^2+ 4\omega_c^2 )\tau_{ee}^2 - 2i\omega\tau_{ee}}
  \Big\{
  \begin{array}{c}
  1- i\omega\tau_{ee}
  \\
  2\omega_c\tau_{ee}
  \end{array}
  ,
\end{equation}
Here $\eta _0 = v_F^2 \tau_{ee}/4$ is the  viscosity
in zero magnetic field.

In Eq.~(\ref{Navier_Stocks_B_with_time_disp_0}) we omit
 the hydrodynamic pressure  term, $-\nabla P /m$, and
the bulk momentum relaxation  term, $-\gamma \mathbf{V}$.
Although a perturbation of the electron density
 $\delta n (\mathbf{r},t)$ leads to a perturbation
  of pressure $\delta P (\mathbf{r},t)$ as well as
  to arising of the electric field $\mathbf{E}_{int}(\mathbf{r},t)$
   due to violation of local charge neutrality,
 the value of $ \nabla \, \delta  P  $ is
 much smaller than $e\mathbf{E}_{int}$ \cite{SI}.
 For weak enough disorder and low temperature,
 the bulk momentum relaxation  term, $-\gamma \mathbf{V}$
  is important only  in the very vicinity of the cyclotron
  resonance  or in the central part
 of very wide samples \cite{SI}.

At  $\omega, \omega_c \gg 1 / \tau_{ee} $ the viscosity coefficients
(\ref{eta zeta ot omega_0}) have a resonance at  $\omega=2 \omega_c$.
 The origin of such  {\em the viscous resonance}
 is  rotation of  the viscous stress tensor
   $\sigma'_{ij} = - m \langle v_i v_j \rangle $
  in magnetic field with is the own   frequency $ 2\omega_c $
($\mathbf{v}$ is the velocity  of  a single electron
 and the triangular brackets  denotes averaging by all electrons).

If the interaction between   2D electrons is strong,
 they must be treated as a Fermi liquid.
 Eqs.~(\ref{cont_0}) and (\ref{Navier_Stocks_B_with_time_disp_0}),
  describing in this case   a fluid consisting
   of the quasiparticles of the Fermi liquid,
 can be derived from the kinetic equation for
 quasiparticles \cite{zeta_F_zidk}.  Herewith the
coefficients $n_0$ and  $\eta_0$
   will contain  not only the characteristics of
    the quasiparticle spectrum, but also the large Landau
     parameters, characterizing the interaction between quasiparticles \cite{zeta_F_zidk}.
     Eigenmodes of a Fermi liquid allow
  hydrodynamic consideration within
  Eqs.~(\ref{cont_0}) and (\ref{Navier_Stocks_B_with_time_disp_0})
provided the following condition: the Landau parameters
in the Fermi liquid  units must be much
greater than unity \cite{future}.
In this case,  the value of the viscosity coefficient $\eta_0$
 is  mainly proportional to the Landau
    parameters, thus the parameter $ v_F = 2 \sqrt{  \eta _0 / \tau_{ee} } $
    is much larger than
  the actual quasi-particle velocity
  at the Fermi sphere, $v_{F0}$: $v_F \gg v_{F0}$.

The electric field $\mathbf{E}( \mathbf{r},t )$
  consists of the two parts:  the field
  of incident radiation $\mathbf{E}_0(t)$
  and the internal field $\mathbf{E}_{int} (\mathbf{r},t)$
induced by the perturbation of the electron density
 $\delta n (\mathbf{r},t) $. We do not
  consider the retardation effects which can be important
   in the region of small wavevectors in some structures
 (see Ref.~\cite{Falko_Khmelnitski,Volkov_Zabolotnykh}).
 Thus we use the electrostatic formula
 $ \mathbf{E} _{int} = - \nabla \delta \varphi $.
 For the structures with a metallic gate located at
 the distance $d$ from the 2D layer
 the electrostatic potential is:
$ \delta \varphi  = (4 \pi e d   /\kappa ) \, \delta n $,
where $\kappa$ is the background dielectric constant.
For the structures without  a  gate the relation
 between $\delta \varphi(\mathbf{r},t)$ and
  $\delta n(\mathbf{r},t)$ is given by the Coulomb law
  with the 2D charge density $e\, \delta n$.

{\em 3. Formation of the linear response.}
There are the two types of waves that can be exited
 in a highly viscous electron fluid: magnetoplasmons
  and   transverse zero magnetosound, the  waves  similar to the
  transverse sound in elastic media.

The magnetoplasmon waves  are described by the one of the two wave solutions
 of Eqs.~(\ref{cont_0}) and (\ref{Navier_Stocks_B_with_time_disp_0})
in the absence of external ac field, $\mathbf{E}_0=0$.
 Neglecting  the relaxation processes,  we obtain
the dispersion law of magnetoplasmons:
$  \omega_{p}(q) =  \sqrt{\omega_c^2 + s^2 q ^2 } $,
where  $ s = 2 \sqrt{  \pi e^2 n_0 d  / m \kappa} $
 is the plasmon velosity which is consider to be much greater than $v_F$.
 Damping coefficient of magnetoplasmons due to viscosity
 is \cite{vis_res}:
  \begin{equation}
\label{Ips_gen_0}
\begin{array}{c}
\displaystyle
\Upsilon_p (q)
 =
   \frac{\omega_c ^2 + \omega_p^2  }{2\omega_p^2    }
  \, q^2 \, \mathrm{Re }  \, \eta_{xx}
 +
   \frac{\omega_c   }{\omega_p }
     \, q^2 \, \mathrm{Im } \, \eta_{xy}
    \:.
 \end{array}
\end{equation}
Here the viscosity coefficients  $\eta_{xx}= \eta_{xx}  (\omega)$
  and $\eta_{xy}= \eta_{xy} (\omega)$     are taken
  at $\omega=\omega_p(q)$. Damping of magnetoplasmons
  due to scattering on disorder in high-quality samples
with very small $\gamma$ is important only in the very vicinity
 of the cyclotron resonance,
  $|\omega - \omega_c|  \lesssim \gamma$ \cite{SI}.

The second wave solution of  Eqs.~(\ref{cont_0})
 and (\ref{Navier_Stocks_B_with_time_disp_0})
 at $\mathbf{E}_0=0$ corresponds to magnetosound, whose
 amplitude $\mathbf{V}_0$ is perpendicular
 to the wavevector $\mathbf{q}$.   Such transverse sound
  is related to an elastic part  of  the shear stress,
  which arises in an inhomogeneous  ac flow of a viscous fluid.
    Within the hydrodynamic approximation,
  the magnetosound dispersion law $\omega_{s}(q)$ and
  its damping coefficient $\Upsilon_s(q)$
  come from  the time dispersion   of viscosity~(\ref{eta zeta ot omega_0}).
  At high frequencies, $\omega_c ,\omega \gg 1/\tau_{ee}$,
   one obtains~\cite{SI}:
  \begin{equation}
  \label{omega_v_0}
  \omega_{s}(q)
  = \sqrt{4\omega_c^2 +\frac{v_F^2 q^2 }{ 4 }}
  \,,
  \qquad
  \Upsilon_s (q)= \frac{4\omega_c^2   + \omega_{s} ^2 }
  {2 \omega_{s} ^2 \tau_{ee} }
  \:.
  \end{equation}
  As it was discussed above, for a strongly non-ideal
 Fermi liquid  the  parameter $v_F$
  is much greater than the actual Fermi
  velocity $v_{F0}$.  Therefore the magnetosound wavelength,
  having the order of magnitude
  $ l_s = l_{\omega} = v_F/\omega  $ at $\omega_c \sim \omega $,
    is much larger than the path that  a quasiparticle passes
  during one period of ac field, $\sim v_{F0} /\omega$.
  Thus we can consider the flows
  with the character spacescales $W$
   as small as  $ v_{F0}/\omega \ll W \ll l_{\omega}  $.
 For our case $s \gg v_F$ the dispersions laws
 $\omega_p(q)$ and $\omega_s(q)$ are shown at Fig.~1(a).

  \begin{figure}[t!]
\centerline{\includegraphics[width=0.9\linewidth]{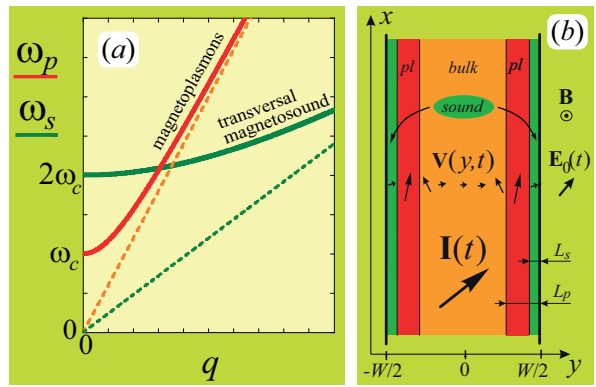}}
\caption{(a) The dispersion laws
 of magnetoplasmons and transverse zero sound.
 The dashed lines demonstrates the dispersion laws in zero magnetic field.
    At $q \to 0$ the viscosity stress tensor
    of the fluid rotates with the frequency  $\omega_s(0) =2 \omega_c$.
  (b)  A long sample in circullary polarized
ac electric field $\mathbf{E_0}(t)$
  and  perpendicular magnetic field $\mathbf{B}$.
    The sample width $W$ is larger than the decay  lengths $L_p$ and $L_s$
    of magnetoplasmons and magnetosound,
    thus the three types of regions are formed: the central region
   where the flow is controlled by the bulk momentum relaxation;
    the near-edge layers with the widths $L_s$ in which
    the flow is governed  by viscosity;
    and the regions $ L_s < |y-W/2| < L_p $
    where the  flow is formed   magnetoplasmons. }
\end{figure}

 Next, we calculate a linear response of the fluid
 on a circularly polarized ac
  electric field  $\mathbf{E}_0(t) = \mathbf{E}_0 e^{-i\omega t} + c.c.$,
  where
$ E_{0,x} = E_0/2 $,  $E_{0,y} = \mp i E_0/2$,
and the signs  ``$-$'' and  ``$+$''
 correspond to the right and the left circular polarizations.
We consider only the structures with a gate,
 believing that the main results will be quantitatively
  the same for ungated structures, while the calculations
  will be far more complex.

We study a flow  in a long sample with rough boundaries,
 at which the zero boundary conditions, $\mathbf{V}(y=\pm W/2) =0 $,
 are fulfilled.  Herewith
the layers formed by the magnetoplasmons and magnetosound are separated
in space  and have the widths $L_p $ and $L_s$  of different orders
of magnitudes   (see Fig.~1(b) and Ref.~\cite{SI}).
 The result for the complex amplitude
 $I=en_0 \int _{- W/2} ^{W/2} \mathbf{V}(y) \, dy$
 of the current $\mathbf{I}(t)$
 in the main order by the parameter $v_F /s \ll 1$ is \cite{SI,comm}:
 \begin{equation}
\label{I_simpl}
\frac{\mathbf{I}}{I_0} =
 \frac{\omega }{  \widetilde{\omega} - \omega_c }
 \Big(\begin{array}{c} i \\ \pm 1 \end{array}\Big)
\pm  \frac{ f_p }{  \widetilde{\omega} - \omega_c }
\Big(\begin{array}{c} i \omega_c  \\ - \omega \end{array} \Big)
 -
 i \, \frac{f_s }{ \omega}
 \Big(\begin{array}{c} 1 \\ 0 \end{array} \Big)
 \:,
 \end{equation}
 where $I_0 = e^2 n_0 E_0 W /(2m\omega)$;
 the first term is the usual Drude contribution,
 which dominates in the central part of wide samples;
 $\widetilde{\omega} = \omega +i \gamma$ takes into account
 a weak scattering on disorder;
 the second and the third terms, being proportional to
 $ f_{p,s} = \tanh(\lambda_{p,s} W/2)/ (\lambda _{p,s} W/2)$,
 describe the plasmonic and the viscoelastic contributions,
 respectively;
   $\lambda_p  = i q_p  - \omega \Upsilon_p / (s^2 q_p)$
 corresponds to  the magnetoplasmons with the decay length
  $L_p=|1/\mathrm{Re} \lambda_p|$, $q_p = \sqrt{\omega^2-\omega_c^2}/s $
  is the magnetoplasmon wavevector; and
 $\lambda_s = -i \omega / \eta_{xx} $ corresponds to the magnetosound
 with the decay length $L_s=|1/\mathrm{Re} \lambda_s|$.

It is noteworthy that the viscoelastic part of the current (\ref{I_simpl})
is independent on the sigh of the circular polarization.

{\em 4. Properties of the linear response.}
The linear response (\ref{I_simpl}) describes, in particular,
 absorbtion    of energy from ac field.
 The dependance of the absorbtion power
 $ \mathcal{W} = 2 \mathrm{Re}(\mathbf{E}_{0}^* \mathbf{I}) $
  on $\omega$ and  $\omega_c$  changes its character
   with changing the  sample width and exhibits the resonances related
 to the Drude, the plasmonic and the viscoelastic parts of  the ac flow.

\begin{figure}[t!]
\centerline{\includegraphics[width=0.9\linewidth]{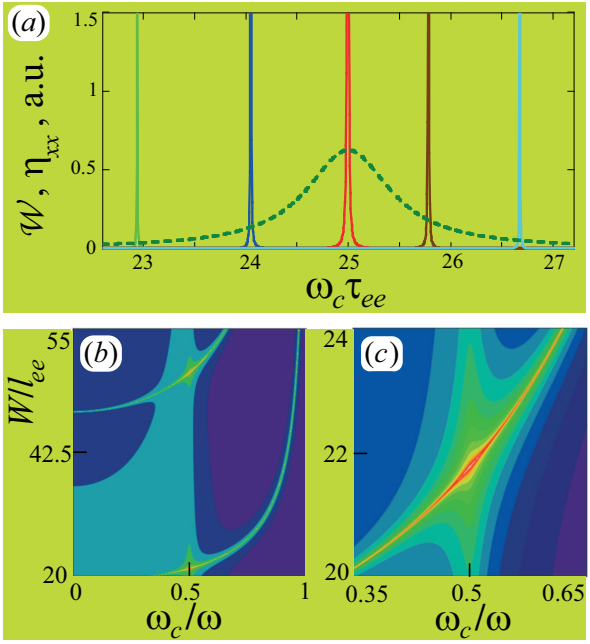}}
\caption{ (a) Absorbtion power $\mathcal{W}$
 as a function of $\omega_c$   at fixed $\omega $
 for the wide samples
  ($W/l_{ee} = 148.5, \, 150.5, \, 152.35, \, 154, \,156 $
 for green, blue, red, brown, and  light-blue solid curves)
 at  $\omega \tau_{ee} = 50$ and $s/v_F = 300$.
 The dependencies $\mathcal{W}(\omega_c)$ exhibit
 the plasmonic resonances,  whose frequencies $\omega_{c,p}$ depend
  on the sample width $W$. Dashed line is
  the dissipative parts of the viscosity coefficients,
  $\mathrm{Re} \, \eta_{xx} =\mathrm{Im} \, \eta_{xy} $.
 The width of plasmonic resonances is maximal
 at the maximum  of $\mathrm{Re} \, \eta_{xx}(\omega_c)$.
   (b,c) Colourmaps  of the function $\mathcal{W} (\omega_c, W)$.
   Maximums of $\mathcal{W}$,
   are located at the intersections of
   the region $\omega_c \approx \omega /2 $
    attributed  the viscous resonance
    by the bent narrow regions attributed  the plasmonic resonances.}
\end{figure}

 In very wide samples, $W \gg L_p $, absorbtion is mainly determined
 by the Drude part of the flow in  the central part
 of the sample [see Fig.~1(b)].  The function
 $\mathcal{W} (\omega_c)$ for a fixed $\omega$
 exhibit the cyclotron resonance at $\omega _c = \omega$.

 In moderately wide samples,  $  l_p  \ll W \ll L_p  $,
the standing waves of magnetoplasmons are formed inside the whole sample
 at ac frequencies above the cyclotron resonance, $\omega > \omega_c$.
 Here $l_p=s/\omega$ is the characteristic magnetoplasmon  wavelength
 at $\omega \sim \omega_c$.
  When the sample width is equal to an integer or a half-integer
 number of magnetoplamon wavelengths,  $    W=  m \pi / q_p(\omega_c)  $
  ($m $ is an integer, $\omega$ is fixed),
  $\mathcal{W} (\omega_c)$  exhibits
   the plasmonic resonances [see Fig.~2(a)].
 The plasmonic resonance  frequency  $\omega_{c,p} (W)$
 can fall in the vicinity of the viscous resonance,
 $ \omega_c = \omega/2 $, or away from it.
  Herewith the width of the plasmonic resonance,
  which is proportional to  the
   damping coefficient due to viscosity  (\ref{Ips_gen_0}),
  resonantly depends on  $ \omega_c - \omega  /2 $ (see Fig.~2).

 \begin{figure}[t!]
\centerline{\includegraphics[width=0.9\linewidth]{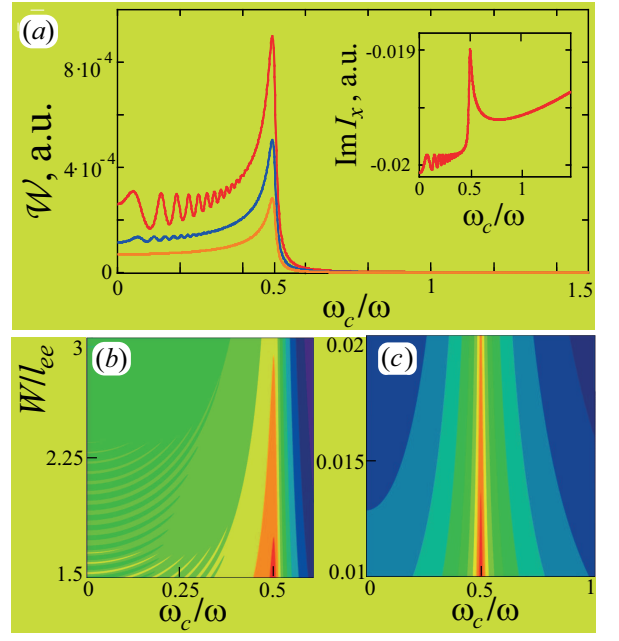}}
\caption{  (a) Absorbtion power $\mathcal{W}$
 as a function of $\omega_c$ at fixed $\omega $ for
  medium and narrow samples widths  $W/l_{ee} = 5.7,\, 3.2,\,  1.8 $
  (yellow, blue,   and red  curves)
  at $\omega \tau_{ee} = 50$ and $s/v_F = 300$.
  Inset presents the imaginary part of the ac current $I_x(\omega_c)$
   for a medium sample ($W/l_{ee} = 1.8$).
  The viscous resonance at $\omega_c = \omega /2 $
  and the magnetosound  resonances
  $\omega_{c,s} = \omega_{c,s} (m) $ are seen.
 (b,c) Colormaps of the functions
 $\mathcal{W} (\omega_c,W)$  and $ R (\omega_c,W)$
  for medium  and   very narrow samples, respectively.
  The profile
   of the viscous resonance is asymmetric for medium and narrow
  samples and symmetric for very narrow samples.
 }
\end{figure}

In medium samples, $   L_s \ll W \ll l_p   $,
at ac frequencies  above the viscous resonance, $\omega > 2\omega_c $
 the flow in the near-edge regions, $ W/2 - |y| \lesssim L_s $,
   have the form of the standing waves of  the transverse zero sound.
  Below the viscous resonance, $  \omega < 2\omega_c $,
  the viscous part of the flow
  is located in the narrower near-edge layers,
   $ W/2 - |y| \lesssim l_{s}$ ($ l_{s} \ll L_s $,
  and has an exponential non-oscillating  profile.
   Due to such change of the type
    of the flow near  $\omega = 2 \omega_c $,
  the viscous resonance arises as an asymmetric peak
  in the dependencies    $\mathcal{W}(\omega_c)$
  and $\mathbf{I} (\omega_c)$ [see Fig.~3(a,b)].
 From Eq.~(\ref{I_simpl}) one obtains
 near the resonance \cite{SI}:
 \begin{equation}
 \label{Visc_res_profile_0}
 \mathcal{W} (\omega_c)  \propto \mathrm{Re}\,
 [  -i  + (2 \omega_c -\omega ) \tau _{ee}]  ^{-1/2}
 \:.
 \end{equation}
 Such peak  is strongly asymmetric
 relative to the point $\omega_c=\omega/2 $ [see Fig.~3(a,b)].

 In narrow samples, $   l_{s} \ll W \ll L_s   $, standing waves
 of magnetosound are formed in the whole sample at $\omega > 2\omega_c $.
 When the sample width becomes equal to an integer
 or a  half-integer number of magnetosound wavelengths,
$    W =  m \pi / q_s(\omega_c)  $,
 the resonances $\omega_{c,s} = \omega_{c,s}  (m)$ related
   to standing magnetosound waves
 appear  in $ \mathcal{W} (\omega_c)$
  ($q_s = 2\sqrt{\omega^2 - 4 \omega_c^2 }/v_F $
  is the magnetosound wavevector).
  Such resonances, shown at  Figs.~2(a,b),
   can be named  the ``magnetosound resonances''.

In maximally narrow samples, $  W \ll l_s  $,
the  velocity $\mathbf{V}(y)$ has a parabolic profile in a whole sample \cite{SI}.
This regime can be regarded as the ac Poiseuille flow in magnetic field.
The viscous  resonance  manifests itself  not in the current,
but in the ac impedance $ Z  = E/I_x $.
 Near the resonance  Eq.~(\ref{I_simpl}) for the real part of the impedance
 $R = \mathrm{Re} \, Z$ yields:
 \begin{equation}
 \label{res_in_imp}
 R (\omega_c) \propto [1 + ( 2 \omega_c  - \omega ) ^2 \tau_{ee} ^2 ]^{-1}
 \,.
\end{equation}
This is the Lorenz profile, symmetric
relative to the point $\omega_c = \omega /2 $ [see Fig.~3(c)].
At $\omega \tau_{ee} \gg 1$ the imaginary part of
the current~(\ref{I_simpl}),
$  \mathrm{Im} \, I_x = I_0 (4 \omega_c^2 - \omega^2  ) W^2  /(3 v_F^2)  $,
does not depend on any relaxation parameter.
 Such result corresponds to
 ac field-induced  oscillations of
  a charged elastic 2D  media, glued  to the edges $y=\pm W/2$.
The inter-particle scattering with the rate $\tau_{ee}$
 is responsible for  damping of this oscillations.

 We conclude that  the viscoelasticity transition
  occurs in the system with decrease of the sample width $W$
 below the magnetosound damping length $L_s$ and
  increase of  ac frequency $\omega $ above $1/\tau_{ee} $.

{\em 5. Manifestation  of the viscous resonance in non-linear effects.}
In high-mobility GaAs quantum wells  bright surprising effects
 were observed at the frequencies $ \omega = 2 \omega_c$
in non-linaer  ac magnetotransport:
a strong peak in photoresistance  \cite{exp_GaAs_ac_1,exp_GaAs_ac_2}
  and  a peculiarity  in the photovoltaic effect \cite{exp_GaAs_ac_3}.
   Below we present  the arguments that
  the viscous resonance is  responsible
for these effects.

It was stressed  in Ref.~\cite{exp_GaAs_ac_1}
  that the strong peak in photoresistance
  and a very well pronounced giant negative magnetoresistance,
 explained in Ref.~\cite{je_visc}
 as a manifestation of forming a viscous flow,
  are observed in the {\em same best-quality} GaAs structures.
   If 2D electrons
   in such structures form a viscous fluid, any ac effect
  must inevitably have  peculiarities
    at the frequency of the viscous resonance.

   We have  demonstrated that the peak of the viscous
   resonance can  be asymmetric as well as symmetric,
    depending on the sample width.    Both these two forms
    of the peak were observed in photoconductivity in
   experiments \cite{exp_GaAs_ac_1,exp_GaAs_ac_2}
    at different conditions .

In Ref.~\cite{exp_GaAs_ac_3} the photovoltage effect was measured
 on several GaAs quantum well structures of different geometries.
 The peculiarity at $\omega = 2 \omega _ c $ was much better observed
 in the  sample having a meander-shaped gate, compared with
 the uniformly-gated and ungated  samples.
 Apparently, the meander-shaped gate
  leads to inhomogeneous perturbations of the electron flow.
  Thus in  the meander-gated sample
  the role of viscosity must be greater than
  in the other samples and  the viscous resonance
   is expected to  be more pronounced, as it was observed.

 In Refs.~\cite{Smet_1,Smet_2,Ganichev_1,Ganichev_2}
  it was shown that  photoresistance of high-mobility
  GaAs  quantum wells is usually almost independent of the sign
  of circular polarization of radiation
 at $\omega_c \lesssim \omega$.   Although the discussed above
 giant peak in photoresistance  was observed up to now only for
the linear polarization of ac field,
a detailed analysis \cite{SI} of the results of
Refs.~\cite{Smet_1,Smet_2,Ganichev_1,Ganichev_2} correlates with
 the statement that  the peak has the hydrodynamic nature.

 The assumption about the strong interaction between quasiparticles
   of the electron Fermi-liquid in the high-mobility GaAs quantum wells
 seems to be fulfilled.  Indeed, the parameter $r_s$ characterizing
  the strength of the Coulomb interaction in  a 2D electron  system
  is about unity for the structures studied in
  Refs.~\cite{exp_GaAs_ac_1,exp_GaAs_ac_2,exp_GaAs_ac_3}.
  In addition, there are experimental evidences
    that the effective mass of 2D electrons
     in the high-mobility GaAs quantum wells
   is strongly renormalized due to the inter-particle
   interaction \cite{book,mass}. Our preliminary analysis \cite{future}
    demonstrates that  the strong interaction between
    quasiparticles in the 2D electron  Fermi liquid
 justify the applicability of hydrodynamics
  for the proper  description of  magnetosonic perturbations in  such
  fluids.

{ 6. Acknowledgements.}
  The authors  thank M.~I.~Dyakonov
 for numerous illuminating discussions of transport phenomena
 in high-mobility two-dimensional electron systems
 which led to the current work, as well as for discussions
  of some of the issues raised in this work;
  I.~V.~Gornyi for kind interest,  discussions, advice,
  and support, as well as for attracting authors' attention
   to some of the references;
    E.~G.~Alekseeva, I.~P.~Alekseeva,  N.~S.~Averkiev, A.~I.~Chugunov,
    A.~P.~Dmitriev,   M.~M.~Glazov,  I.~V.~Krainov, A.~N.~Poddubny,
 P.~S.~Shternin, D.~S.~Svinkin, V.~A.~Volkov, A.~A.~Zyuzin, and A.~Yu.~Zyuzin
   for valuable discussions, advice, and support;
and D.~G.~Polyakov for reading the preprint of the manuscript
 and the critical remarks that helped to  substantially improve it.

 The part of this work devoted to calculation
  of the linear response of the viscous electron fluid
  (Section ``Formation of the linear response'')
 was supported by
 the Russian Science Foundation (Grant No.  17-12-01182);
 the part of this work
 devoted to the study of the ways of manifestation of the viscous resonance
in the viscous electron fluid
 (Section ``Properties of the linear response'')
   was supported by the grant of the Basis Foundation  (Grant
No. 17-14-414-1);
 the part of this work devoted to analysis
 of applicability of the hydrodynamic model
 of strongly non-ideal electron fluid
 for real high-mobility GaAs quantum wells
 (Sections ``Ac hydrodynamics of 2D electron  Fermi gas
and strongly non-ideal Fermi liquid''
 and ``Manifestation  of the viscous resonance  in non-linear effects'')
 was supported by
 the Russian Science Foundation (Grant No.  18-72-10111).

\appendix

\newpage

\section{Supplemental}
\section{1. Calculation of linear response
of 2D electron  fluid on ac electric field}
In this section we  present a calculation
of the  flow  of a highly viscous 2D electron fluid
 in a long sample with rough boundaries.

  In the regime linear by external ac electric filed,
   the continuity and the Navier-Stocks equation
   for an electron fluid in magnetic field,
  taking into account bulk momentum relaxation, are \cite{vis_res_}:
\begin{equation}
\label{cont}
-i\omega \, \delta  n + n_0\mathrm{div }  \mathbf{V} = 0
 \:,
 \end{equation}
 and
 \begin{equation}
\label{Navier_Stocks_B_with_time_disp}
 \begin{array}{c}
  \displaystyle
-i\omega  \mathbf{V}
=
 \frac{e}{m}\mathbf{E}(\mathbf{r},\omega)
  + \omega_c \left[\mathbf{V} \times \mathbf{e}_z \right]
    - \gamma\mathbf{V}
 -\frac{\nabla P }{m} +
\\
\\
\displaystyle
+ \eta_{xx} (\omega) \, \Delta \mathbf{V}+  \eta _{xy} (\omega)
  \left[  \Delta \mathbf{V} \times \mathbf{e}_z \right]
  \end{array}\:,
\end{equation}
where   $\mathbf{E}(\mathbf{r},\omega) $,
 $ \delta n = \delta n(\mathbf{r},\omega)  $ and
$\mathbf{V} = \mathbf{V} (\mathbf{r},\omega) $  are the complex amplitudes
of the harmonics of the electric field $ \mathbf{E} (\mathbf{r},t )  $,
 the particle density $ \delta n(\mathbf{r},t )  $,
 and the hydrodynamic velocity   $\mathbf{V} (\mathbf{r},t ) $;
 $n_0$ is the unperturbed electron  density;
 $e<0$ is the electron charge; $m$ is the electron mass;
  $\gamma$ is the rate of momentum relaxation in bulk due to the
 electron scattering  on disorder or phonons;
  $P$ is the pressure in the electron gas; and the ac
  viscosity coefficients $\eta_{xx} (\omega)$ and $\eta_{xy} (\omega)$
 depend on magnetic field and frequency as \cite{vis_res_}:
\begin{equation}
\label{eta zeta ot omega}
\begin{array}{c}
 \displaystyle
\eta_{xx} (\omega) =\eta_0 \frac{1- i\omega\tau_{ee}}
 {1+(-\omega^2+ 4\omega_c^2 )\tau_{ee}^2 - 2i\omega\tau_{ee}}
  \:,
\\
 \\
\displaystyle
\eta_{xy} (\omega) =\eta_0 \frac{2\omega_c\tau_{ee}}
 {1+(-\omega^2+ 4\omega_c^2 )\tau_{ee}^2 - 2i\omega\tau_{ee}}
 \:.
\end{array}
\end{equation}
Here $\eta _0 = v_F^2 \tau_{ee}/4$ is the 2D electron viscosity
in the absence of magnetic field;  $\tau_{ee}$ is the time
 of relaxation of the second harmonics (by the electron velocity angle)
 of the electron distribution function;
 and $v_F$ is the   Fermi velocity  $v_{F0}$
 for the case of a weak electron-electron interaction
 and the parameter determined as $ v_F = 2 \sqrt{ \eta/\tau_{ee}}$
 for a strong inter-quasiparticle interaction.
 In the last case, $v_F$ depends on
 the parameters of the Landau function describing the interaction of quasiparticles
and is much greater than the Fermi velocity  $v_{F0}$.

The electric field $\mathbf{E}( \mathbf{r},t )$
 consists of the two parts:  external circularly polarized
  electric field $\mathbf{E}_0(t)$
  and the internal electric field $\mathbf{E}_{int} (\mathbf{r},t)$
induced by the perturbation of
the 2D electron density $\delta n (\mathbf{r},t) $.
In this work we do not consider the retardation effects
 which can be important in the region
 of small wavevectors in some structures
 (see Ref.~\cite{Falko_Khmelnitski_,Volkov_Zabolotnykh_}).
 When we can neglect the retardation effects,
 we just have the electrostatic formula
 $ \mathbf{E} _{int} = - \nabla \delta \varphi $,
 where the electrostatic potential
 $\delta \varphi $ is related to $\delta n$.
 For the structures with a metallic gate located at
 the distance $d$ from the 2D layer we have:
\begin{equation}
\label{el}
\delta \varphi  = \frac{ 4 \pi e d   }{\kappa } \, \delta n
\:,
\end{equation}
where $\kappa$ is the background dielectric constant.
For the structures without  a  gate the relation
 between $\delta \varphi(\mathbf{r},t)$ and $\delta n(\mathbf{r},t)$
 is given just by the Coulomb law with the charge density
 $\varrho(\mathbf{r},z,t) =  e \,  \delta n (\mathbf{r}, t ) \, \delta(z) $, where
 $\delta(z)$ is the Delta-function depicting
 the position of the 2D layer:
\begin{equation}
\label{el_2}
\delta \varphi (\mathbf{r},t)  = e \int d^2\mathbf{r}'
 \, \frac{\delta n (\mathbf{r}',t)}{|\mathbf{r}-\mathbf{r}'|}
\:.
\end{equation}

The ratio of the terms $-\nabla P /m $ and $e\mathbf{E}_{int} /m $
 in Eq.~(\ref{Navier_Stocks_B_with_time_disp}) is estimated
as $a_B /d $ for the structures with a gate
and as $a_B q    $ for the ungated structures,
 where $a_B$ is the Bohr radius and $q$ is the character wavevector.
Both these values must be much smaller than
 unity when the 2D electrostatic equations
 are applicable.

 Neglecting  the relaxation processes and in the absence
  of the external electric field,
 $\mathbf{E}_0=0$,  we obtain from
Eqs.~(\ref{cont}), (\ref{Navier_Stocks_B_with_time_disp}),
 and (\ref{el}) the  usual formula
 for the dispersion law of magnetoplasmons.
 For the gated structures it is:
 \begin{equation}
 \label{omega_p}
  \omega_{p}(q) =  \sqrt{\omega_c^2 + s^2 q ^2 }
  \:,
  \end{equation}
where  $ s = \sqrt{ 4 \pi e^2 n_0 d  / m \kappa} $.
The second term under the square root, $s^2q^2$,
 is the squared plasmon frequency in the absence
 of magnetic field. For the ungated structure
it changes on  $ 2 \pi e^2 n_0   q/ m \kappa$.

 We calculate a linear response of the electron fluid
 on the circularly polarized homogeneous ac
  electric field
   $\mathbf{E}_0(t) = \mathbf{E}_0 e^{-i\omega t} + c.c.$,
\begin{equation}
 \label{E_0}
\mathbf{E}_0 = \frac{E_0}{2}
 \left(
 \begin{array}{c}
 1
 \\
 \mp i
 \end{array}
 \right)
 \:,
\end{equation}
where  the signs ``$-$'' and  ``$+$''
 correspond to the right and the left circular polarizations
 of ac field.

We confine ourselves by consideration only the case of gated structures.
 We believe that the main physical results will be quantitatively
  the same for ungated structures, while the calculations
  will be far more complex for ungated structures
  due to non-local character of electrostatics
   in  the ungated case [see Eq.~(\ref{el_2})].

Let us consider a high-frequency flow in a long sample,
 $-L/2 \ll x \ll L/2 $, $-W/2 \ll y \ll W/2 $,  $L \gg W$,
  with rough longitudinal boundaries.
 This is the simplest ``minimal'' model for studying
 viscous transport in the confined geometry.
  The velocity field $\mathbf{V} (\mathbf{r},t)$  have the form:
$\mathbf{V} (\mathbf{r},t) = \mathbf{V} (y) e^{-i \omega t} + c.c. $,
 and the analogous  formulas take place
for $\delta n (\mathbf{r},t) $ and $\delta U (\mathbf{r},t) $.
    Substitution of $\delta n (y)$ and $\delta \varphi (y)$
  from the continuity and the electrostatic equations (\ref{cont})
   and (\ref{el})  to the Navier-Stocks equation
    (\ref{Navier_Stocks_B_with_time_disp}) yields the closed equation for
  the complex amplitude of the velocity $\mathbf{V}(y)$:
\begin{equation}
\label{matr_eq}
\left(
\begin{array}{cc}
\displaystyle
i\widetilde{\omega}  + \eta  \, \partial^2   &   -(\omega_c + \overline{\eta}  \, \partial^2 )
\\
\\
\displaystyle
\omega_c + \overline{\eta} \,  \partial^2 &  i\widetilde{\omega }
 + \Big(\eta + i \, \frac{ \displaystyle s^2}
 { \displaystyle \omega} \Big) \, \partial^2
\end{array}
\right) \mathbf{V}
  =
  - \frac{e \mathbf{E}_0}{m}
 \:,
\end{equation}
where we introduce the simplified notations:
$\partial = \partial / \partial y$,
$\widetilde{\omega} = \omega  + i \gamma $,
$\eta = \eta_{xx} (\omega)$,
and  $\overline{\eta} = \eta_{xy} (\omega)$.
 Equation (\ref{matr_eq}) for the rough sample edges
 should be supplemented by the diffusive boundary conditions:
\begin{equation}
 \label{bound_eq}
\mathbf{V}|_{y=\pm W/2} = 0
 \:.
\end{equation}

We suppose  that  the rate of electron scattering
on disorder,  $\gamma$, is the smallest value among
all  other frequencies and rates of the problem: $\gamma \to 0$.
The equation for the eigenvalues in  this case is:
\begin{equation}
 \label{lambda_eq}
\begin{array}{c}
\displaystyle
\Big(i \frac{s^2}{\omega} \eta + \eta^2
 +\overline{\eta}^2 \Big)
 \lambda ^4 + (-s^2 + 2i\omega \eta
 + 2 \omega_c \overline{\eta}) \, \lambda^2 +
 \\
 \\
+ \, \omega_c^2 -\widetilde{\omega}^2  = 0
 \:.
\end{array}
\end{equation}
The scattering rate $\gamma$ should be kept only
in the last term as the other terms have nonzero
real as well as imaginary parts
 already at $\gamma = 0$.

The most transparent for analysis is the regime,
 when the spacescales corresponding to
the two roots $\lambda_1$ and $\lambda_2$
  of Eq.~(\ref{lambda_eq}) have different orders of magnitudes:
  $ | \mathrm{Re} \lambda_1 | , \,  | \mathrm{Im} \lambda_1 |
   \ll
    | \mathrm{Re} \lambda_2 | ,  \,  | \mathrm{Im} \lambda_2 |  $.
  We will show below    that this  case is realized when
\begin{equation}
\label{crit}
 \frac{s }{v_F}  \gg  \omega \tau_{ee} \gg 1
 \:.
 \end{equation}
 If this inequality is satisfied, calculations based on
 Eqs.~(\ref{eta zeta ot omega}) and ~(\ref{lambda_eq})
 leads to:
\begin{equation}
 \label{lambdas}
\lambda_1 ^2 =
  \frac{ \omega_c^2 -\widetilde{\omega} ^2   }{s^2}
  \: ( \, 1 + i \, u \, )
  \:, \quad
\lambda_2 ^2 =
-i \, \frac{\omega}{\eta} \, ( \, 1 - i \, w \, )
\:,
\end{equation}
where the small parameters $u$ and $w$, $|u|,|w| \ll 1$, are:
\[
 u=
 \frac{ ( \omega^2 + \omega_c^2 )\, \eta -2i
   \omega \omega_c \,  \overline{\eta} } {\omega  s^2 }
   \:, \qquad
 w=
 \frac{ ( \omega_c  \eta -  i \omega \overline{\eta } )^2 }{\omega \eta s^2}
 \:.
\]

 The first eigenvalue $\lambda_1 $ describes the  character length
  of the part of the response related to perturbation of charge,
  that is to magnetoplasmons.
     The corresponding to $\lambda_1 $
   damping coefficient $\Upsilon _p(q) $
  outside the cyclotron resonance, $ |\omega-\omega_c| \gg \gamma$,
   was calculated in Ref.~\cite{vis_res_}:
  \begin{equation}
\label{Ips_gen}
\begin{array}{c}
\displaystyle
\Upsilon_p (q)
 =
   \frac{\omega_c ^2 + \omega_p^2  }{2\omega_p^2    }
  \, q^2 \, \mathrm{Re }  \, \eta
 +
   \frac{\omega_c   }{\omega_p }
     \, q^2 \, \mathrm{Im } \, \overline{\eta}
    \:,
 \end{array}
\end{equation}
where the viscosity coefficients
$ \eta = \eta (\omega)$ and $\overline{\eta} = \overline{\eta} (\omega)$
are taken at  the magnetoplasmon frequency $ \omega = \omega_p(q) $.
Damping of magnetoplasmons due to scattering on
disorder is important only
in the vicinity of the cyclotron resonance,
$|\omega - \omega_c|  \lesssim \gamma$
 [see Eq.~(\ref{lambdas})].

The second eigenvalue $\lambda_2 $ corresponds
   to the standing waves of the transverse  zero sound.
Such waves exists in strongly non-ideal Fermi liquid
  at the frequencies $\omega \gg 1/\tau_{ee}$ \cite{LP_}.
  They are similar to the transverse sound in
  highly viscous amorphous media, such as glasses, plastics, etc.
  The transverse zero sound is a characteristic phenomena to distinguish
  weakly (gas-like) and strongly (honey-like) viscous media:
     it is absent in the first and exist in the last.
   When the non-ideality of the Fermi liquid is very high,
 the longitudinal and the transverse   zero sounds can be considered
  within hydrodynamics \cite{future_}.
  The  electron-electron collisions, controlling relaxation
 of the viscous stress tensor,
 determine the damping coefficient  of the transverse sound.

  The dispersion law $\omega_{s}(q)$
 and the damping coefficient $\Upsilon_s(q)$
  of the transverse zero sound
 come from the second of Eq.~(\ref{lambdas}):
 \begin{equation}
 \lambda_2 ^2
 \approx
 - i \frac{ \omega }{\eta(\omega,\omega_c) }
 \:,
 \end{equation}
  if we put  $\lambda_2 = iq$ ($q$ is a real wavevector)
  and let $\omega $ have an imaginary part,
 proportional to the magnetosound damping coefficient $\Upsilon_s$.
  At high frequencies, $\omega_c ,\omega \gg 1/\tau_{ee}$, and
   far from the viscous resonance,
    $| \omega - 2 \omega_c | \gg 1/\tau_{ee}$,
   we obtain from Eq.~(\ref{lambdas}):
  \begin{equation}
  \label{omega_v}
  \omega_{s}(q)
  = \sqrt{4\omega_c^2 +\frac{v_F^2 q^2 }{ 4 }}
  \:,
  \qquad
  \Upsilon_s (q)= \frac{4\omega_c^2   + \omega_{s} ^2 }
  {2 \omega_{s} ^2 \tau_{ee} }
  \:.
  \end{equation}
 As we discussed above, for a strongly non-ideal Fermi liquid
 $v_F$ is a parameter
that enters  in the zero-field viscosity coefficient
 as $\eta _0  = v_F^2 \tau_{ee}/4$ and  is much greater than
  the actual Fermi velocity $v_{F0}$ \cite{future_}.
  As a consequence, the minimal character length
  in the present theory, $l_{\omega}=v_F/\omega$,
  is much larger than the length of the path that  a quasiparticle passes
  during one period of variation of ac field,
   $\sim v_{F0} /\omega$. Herewith we consider that $s \gg v_F$.
  Thus we are able to consider the flows $\mathbf{V}(y)$
  having the character spacescales $W$ as small as
  $ v_{F0}/\omega \ll W \ll l_{\omega}  $.

In Fig.~1(a) of the main text  we have presented the dispersion laws
 of the  magnetoplasmons and the transverse zero sound.
In this work we do not study the detailed structure of waves
near  the  interaction point of
 the  magnetoplasmons and the transverse zero sound  dispersion laws
  $\omega_p(q)$ and $\omega_s(q)$.

The straightforward calculations on base of Eqs.~(\ref{matr_eq}),
  (\ref{bound_eq}), and (\ref{lambdas})
  lead to the following result for the distribution of the velocity:
\begin{equation}
 \label{V_x}
   \mathbf{V} (y)=\frac{eE_0}{2m}
\left[
\mathbf{A}_0
 + h(\lambda_1,y)\, \mathbf{A}_1
+h(\lambda_2,y)\, \mathbf{A}_2
\right]
 \:,
  \end{equation}
where
\[
\mathbf{A}_0=
\frac{1}{ \widetilde{\omega } \pm \omega_c }
\left(
 \begin{array}{c}
 i
 \\
 \pm 1
 \end{array}
\right)
\:,
\quad
h(\lambda,y) = \frac{\cosh(\lambda y )}{\cosh(\lambda W/2 ) }
\:,
\]
\[
\mathbf{A}_1 =
 \pm  \frac{ 1}{ \omega  \, ( \widetilde{\omega} \pm \omega_c) }
\left(
 \begin{array}{c}
  i\omega_c
   \\
 - \omega
 \end{array}
\right)
+
 \frac{\varepsilon    }{\omega^2}
\left(
 \begin{array}{c}
  - 2 \omega_c   \pm  \omega
   \\
  - i \omega
 \end{array}
\right)
       \:,
\]
\[
\mathbf{A}_2 =
  \frac{ i }{ \omega   }
\left(
 \begin{array}{c}
  -  1
   \\
 \varepsilon
 \end{array}
\right)
\left[ 1
+
  \frac{i \varepsilon    }{\omega}
\, (
 \,  2 \omega_c   \mp  \omega
\, )
 \right]      \:,
\]
and $ \varepsilon = (\omega_c \eta -  i \omega \overline{\eta} )/ s^2 $.

It is seen from Eq.~(\ref{V_x})  that the viscoelastic  part
of the linear response is responsible for
 forming the flow in the very vicinities of the sample edges
 with the widths of the order of $L_s = 1/ |\mathrm{Re} \lambda_2|$ in the case
 $W \gg L_s $ [see Fig.~1(b) in the main text]
 or in the whole sample for very narrow samples, $W \ll L_s $.
  The plasmonic part of the linear response  governs the flow
  in the wider near-edge layers with the widths
  of the order of $ L_p = 1/| \mathrm{Re} \lambda_1| $
  in the very wide samples, $W \gg L_p $ [see Fig.~1(b) in the main text],
  or in the central part  of the samples
  with the widths in the interval  $ L_s \ll W \ll L_p$.
 In the central part of the very wide samples, $W \gg L_p $,
  the flow is Ohmic: $\mathbf{V} (y) =  eE_0 \mathbf{A} _0 / (2m) $
   [see Fig.~1(b) in the main text].
  It  is controlled just by motion of individual electrons
 in external electric and magnetic fields
 and scattering on disorder.

For the complex amplitude
 \[
 \mathbf{I} = e n_0 \int _{-W/2}^{W/2} dy \: \mathbf{V}(y)
 \]
of the total electric current,
  $\mathbf{I}(t) =  \mathbf{I} e^ {-i \omega t } + c.c.$,
   we easily get from Eq.~(\ref{V_x}):
\begin{equation}
 \label{I_x_gen}
    \mathbf{I} =\frac{e^2 n_0 E_0W}{2m}
\left[
\mathbf{A}_0
 + f(\lambda_1)\, \mathbf{A}_1
+ f(\lambda_2)\, \mathbf{A}_2
\right]
\:,
\end{equation}
 where
\[
f(\lambda) = \frac{ \tanh (\lambda W/2) }{ \lambda W/2}
\:.
\]

Below  in this section we perform analytical estimations  of the eigenvalues
 $\lambda_1$ and $\lambda_2$   for the most interesting
  case of the large frequencies $\omega$ and $\omega_c$ having
   the same order of magnitude, being mach greater than $ 1/\tau_{ee}$,
  and lying far or near the cyclotron  and the viscous
   resonances.

 For the frequencies $\omega $  above the cyclotron resonance,
$ \omega -  \omega_c  \gg \gamma$,
the imaginary part of the plasmonic eigenvalue $\lambda_1$ is:
\begin{equation}
 \label{pl_wavelength}
\mathrm{Im}  \,  \lambda_1 \approx
 \frac{\sqrt{\omega^2 -\omega_c^2 }}{s}
  \sim \frac{1}{ l_p  }
 \:.
\end{equation}
Here and below we present, if it is possible, we present only
the positive values of the imaginary
and the real parts of the eigenvalues $\lambda_{1,2}$.
 The value  of $\mathrm{Im}  \lambda_1 $ in Eq.~(\ref{pl_wavelength})
 is much grater than the real part $\lambda_1 $, which is:
\begin{equation}
  \label{pl_length_of_decay_gen}
\mathrm{Re} \, \lambda_1 \approx
\frac{1}{l_p  \, s^2    }
\Big( \, \frac{\omega^2 + \omega^2_c }{ 2 \omega}
\, \mathrm{Re} \, \eta
+  \omega_c    \, \mathrm{Im} \,
 \overline{\eta} \, \Big)
 \:.
\end{equation}
The last formula corresponds to the damping coefficient
of magnetoplasmons due to viscosity (\ref{Ips_gen}).
 The values $\mathrm{Re} \, \eta $
 and $ \mathrm{Im} \, \overline{\eta} $,  in their turn,
  strongly depend on the closeness of the frequencies
   $\omega$ and $\omega_c$ to the viscous resonance.
    For the frequencies far and in vicinity of the viscous resonance
     we obtain the two possible magnitudes of
 $\mathrm{Re} \, \lambda_1$:
 \begin{equation}
 \label{pl_length_of_decay_exact}
\mathrm{Re} \, \lambda_1   \sim
 \frac{1  }{l_p} \frac{v_F^2}{s^2}
 \left\{
 \begin{array}{l}
 1/( \omega \tau_{ee} )\, , \;\; |\omega - 2 \omega_c| \gg 1/\tau_{ee}
 \\
 \\
 \omega \tau_{ee} \, , \;\; |\omega - 2 \omega_c| \lesssim 1/\tau_{ee}
 \end{array}
 \right.
 .
  \end{equation}
We see from Eqs.~(\ref{pl_wavelength}) and
(\ref{pl_length_of_decay_exact})   that indeed
 $\mathrm{Im} \,  \lambda_1  \gg \mathrm{Re} \,  \lambda_1  $, if
the condition (\ref{crit}) is fulfilled.

 Accordingly to Eqs.~(\ref{pl_wavelength}) and (\ref{pl_length_of_decay_exact}),
  above the cyclotron resonance
  the velocity profile $\mathbf{V}(y) $ is a standing magnetoplasmon wave
 with the wavelength $l_p \sim s/\omega$ and  the length of decay
 $ L_{p}^{out} \sim (s/v_F)^2  \, s\tau_{ee}  $ outside the viscous resonance
  or $ L_{p}^{in} \sim (s/v_F)^2 [s/(\omega^2 \tau_{ee})]$
   in the vicinity of the viscous resonance.

Below the cyclotron resonance,
$ \omega_c -  \omega \gg    \gamma$,
the real part of the plasmonic eigenvalue $\lambda_1$
 is much greater than its imaginary part:
\begin{equation}
 \label{pl_length_of_imagnary_plasmon}
\lambda_1 \approx
\frac{\sqrt{\omega_c^2 -\omega^2 }}{s}
 \sim \frac{1}{ l_p }\:.
\end{equation}
Thus the corresponding distribution of $\mathbf{V}(y)$
 has the non-oscillating exponential
 profile with the character length $l_p = s/\omega$.

In the vicinity of the cyclotron resonance,
   $ | \omega -  \omega_c | \lesssim \gamma$,
 the real as well as the imaginary parts of
the plasmonic eigenvalue are controlled
by scattering on disorder:
\begin{equation}
 \label{disp_on_CR}
\lambda_1  \approx
\frac{\sqrt{ \omega _c ^2 -\omega^2  - 2i\gamma \omega } } {s}
 \:.
\end{equation}
Such $\lambda_1 $ corresponds to oscillations and
 exponential decay of $\mathbf{V}(y)$ with the same period
 and the decay length $ \widetilde{L}_p \sim  s/\sqrt{\gamma \omega}$.
At very weak scattering on disorder, $\gamma \to 0 $,
 the lengthscale $ \widetilde{L}_p  $
is far greater than the lengths $l_p$, $L_{p}^{in}$ and $L_{p}^{out}$
of decay of the plasmonic contribution
  outside the cyclotron resonance.

The eigenvalue $\lambda _2$ depends only on the closeness
of the frequencies $\omega$ and  $\omega_c$ to the viscous resonance.
Above the viscous resonance,
$ \omega - 2 \omega_c \gg 1/\tau_{ee}$,
we have:
\begin{equation}
 \label{lambda_2_waves_Im}
\mathrm{Im} \,  \lambda_2  \approx
 \frac{  2 \sqrt{ \omega^2 - 4 \omega_c^2}  }{ v_F}
 \sim \frac{1}{l_{\omega}  }
\end{equation}
and
\begin{equation}
 \label{lambda_2_waves_Re}
\mathrm{Re} \,  \lambda_2  \approx
 \frac{ 2 \, ( 4 \omega_c^2 +\omega^2 )}
 {v_F^2 \,  \omega \, \tau_{ee} \,\mathrm{Im} \,  \lambda_2  }
\sim
\frac{  1  }{ l_{ee} }
 \:.
\end{equation}
Here $l_{ee} = v_F \tau_{ee}$ is the parameter
that is not the real scattering length of quasiparticles, but
 is the lengthscale  related
 with the viscosity coefficient $\eta_0$
  as  $ l_{ee} = v_F\tau_{ee}  2 \sqrt{\eta_0  \tau_{ee}  }$,
  according to definition of $v_F$ (see above). We see that
Eqs.~(\ref{lambda_2_waves_Im}) and (\ref{lambda_2_waves_Re})
correspond to the dispersion law and the damping coefficient
of the transverse zero sound (\ref{omega_v}). As
$\mathrm{Im} \,  \lambda_2  \gg \mathrm{Re} \,  \lambda_2  $,
  above the viscous resonance, $\omega - 2 \omega_c \gg  1 / \tau_{ee}$,
  the eigenvalue $\lambda _2$
   corresponds to the standing waves of the  transverse zero sound
 with the wavelength of the order of $ l_s  \equiv  l_{\omega}$ and  the length of decay
 $ L_s \sim l_{ee}$.

 Below the viscous resonance,
$ 2 \omega_c  - \omega \ll  1/\tau_{ee} $,
the eigenvalue $\lambda_2$ is mainly real:
\begin{equation}
\label{lambda_2_decay}
\lambda_2  \approx
   \frac{  2 \sqrt{   4 \omega_c^2 - \omega^2}  }
   { v_F}  \sim \frac{1}{l_{\omega}  }
 \:.
\end{equation}
 In this case the viscoelastic part of the response
  has the form of  the exponential increase of
  $\mathbf{V}(y)$  from the edges into the bulk with
   the character length $l_{\omega}$.

In the vicinity of the viscous resonance,
$| \omega - 2 \omega_c | \lesssim 1/\tau_{ee}$,
 we obtain from Eq.~(\ref{lambdas})
 the following result:
\begin{equation}
 \label{lambda_2_in_visc_res}
 \lambda_2   \approx
 \frac{ 2 \sqrt{4\omega_c^2 -\omega^2 - 4i \omega/\tau_{ee} }  }
 {   v_F }
 \:.
\end{equation}
 In this case, the eigenvalue $\lambda_2$,
  as the eigenvalue  $\lambda_1 $ in Eq.~(\ref{disp_on_CR})
  near the cyclotron resonance,  corresponds to exponential decay
  and oscillations with the same character lengths
  $\sim  \sqrt{l_{\omega } l_{ee} } $.
  The last value  is much smaller than decay length, $L_s \sim l_{ee}$,
  of magnetosound above the viscous resonance
  and is much greater than the width of the near-edge regions,
  $l_s $,   below the viscous resonance.

We see that the behavior of the eigenvalue $\lambda_2   $
near and far from the viscous resonance is similar
 to the behavior of the eigenvalue $\lambda_1   $
 near and far from the cyclotron resonance.

 As it is seen from Eqs.~(\ref{pl_wavelength})-(\ref{lambda_2_waves_Re}),
    the introduced above condition (\ref{crit})
  guarantees that  $L_p,l_p \gg L_s,l_s$.
 Thus the  plasmonic and the viscoelastic parts of
  the linear response  are located in the well-defined
layers with the widths of the different
 orders of magnitude [see Fig.~1(b) of the main text].

  An analytical  calculation of $\mathbf{V}(y)$
  in a general case  when the condition (\ref{crit})
  is failed and  the  character lengths of the plasmonic
  and the viscoelastic part of the linear response
  can intersect each other is very cumbersome.
  In this case,   it is reasonable
  to use  numerical calculations to obtain the resulting
dependencies of $\mathbf{V}(y)$ and $\mathbf{I}$
on $\omega$ and $\omega_c$. However, as it is seen from
 Eq.~(\ref{lambda_eq}), the values of $\mathbf{V}(y)$ and
  $\mathbf{I}$ will still exhibit
 the cyclotron and the viscous resonances.

\section{2. Analysis of the linear response}
In this section we perform an analysis
of the properties of the linear response
of a 2D highly viscous electron fluid
 in a long sample.

The linear response $\mathbf{V}(y,t)$ calculated in the above section
 describes, in particular,
 absorbtion of energy from the ac external field $\mathbf{E}_0(t)$.
The absorbtion power is:
 \begin{equation}
 \label{W}
\mathcal{W} = 2 \mathrm{Re}(\mathbf{E}_{0}^* \mathbf{I}) =
 E_0 \mathrm{Re} (I_x \pm i I_y)
  \:.
\end{equation}
According to the above discussion, depending on
the sample width $W$ and the frequencies $\omega$ and $\omega_c$,
   this value can be mainly determined by magnetoplasmons
   or by  viscosity. We calculated the dependencies
   $\mathcal{W}(\omega_c)$ by Eqs.~(\ref{I_x_gen})
 and (\ref{W}) at a fixed ac frequency $\omega$
 for different  sample widths $W$
 (see Fig.~4 and its descriptions below).

\begin{figure}[t!]
\centerline{\includegraphics[width=1.1\linewidth]{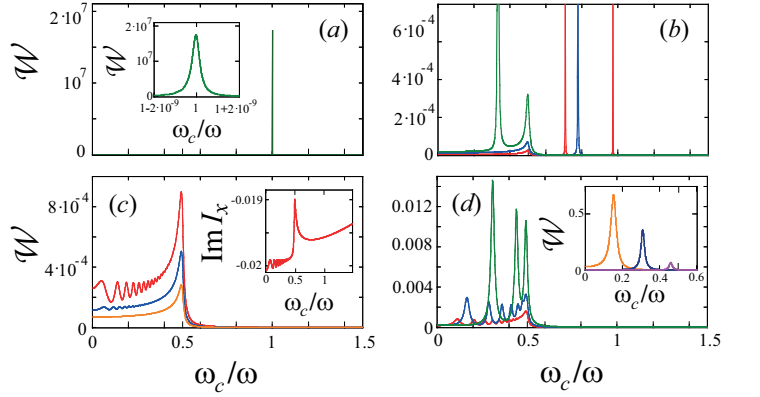}}
\caption{ The absorbtion power $\mathcal{W}$
 in units of $\mathcal{W}_0= e^2 n_0 W E_0^2 \tau_{ee} /(2m)$
 as a function of the cyclotron frequency $\omega_c$
  at a fixed ac frequency $\omega $
  for the very wide, wide, medium, and narrow samples.
  All curves are plotted for the following  parameters:
$\omega \tau_{ee} = 50$, $s/v_F = 300$, and
 $\gamma \tau_{ee} = 10^{-7}$.
Panel (a) demonstrates $\mathcal{W}(\omega_c)$
for the very wide samples with $W/l_{ee} = 10^6$,
   that corresponds to
   $W/l_p \approx 2\cdot 10^5 $,
   $W/L_{p } ^{out} \approx 0.04  $,
   and  $W/L_{p} ^{in} \approx  10  $. Panel (b) corresponds to
 the wide samples with $W/l_{ee} =80,\, 30, \, 20 $
 (red, blue, and green curves);
  these values of  $W/\l_{ee}$ correspond to
  $ 2 \pi W/l_p \approx  5.0,\, 1.2 ,\, 0.9$.
   Panel (c) shows $\mathcal{W}(\omega_c)$ for
 the medium samples with $W/l_{ee} = 5.7,\, 3.2,\,  1.8 $
 and $W/l_{\omega} \approx 280,\, 160,\,90  $
  (yellow, blue,   and red  curves).
 Panel (d) corresponds to the narrow samples
 with $W/l_{ee} =1,\, 0.5,\, 0.2 $ and
  $W/l_{\omega} = 50,\, 25,\, 10   $
  (red, blue, and green curves).
 Inset on panel (a) shows $\mathcal{W}(\omega_c)$
 for the very wide sample with $W/l_{ee} = 10^6$
 in large  scale by the horizontal axis.
 Inset on panel (c) presents the imaginary part
  of the ac current $I_x(\omega_c)$
 in units of $I_{x0}= e^2 n_0 W E_0 \tau_{ee} /(2m)$
 for the case of the medium sample ($W/l_{ee} = 1.8$).
Inset on panel (d) presents
the absorbtion power $\mathcal{W} (\omega_c)$
 in units of $\mathcal{W}_0$
 for the very narrow samples
 ($W/l_{ee} = 0.08,\, 0.04,\, 0.033 $ and
 $W/l_{\omega} \approx 0.4, \, 0.2,\, 0.16$
  for purple, blue, and yellow curves);
  one can see disappearance  of the last transverse
   sound resonances with decrease of $W$.
 }
\end{figure}

{ \em a) Wide samples: the behaviour of the plasmonic contribution. }

 In very wide samples:
  \begin{equation}
   \label{very_wide}
 W \gg L_p \gg  l_p
 \:,
  \end{equation}
  where $L_p = 1/\mathrm{Re}\lambda_1   $ and $l_p =s/\omega$,
the plasmonic perturbation at the frequencies  above the cyclotron resonance,
 $\omega - \omega_c \gg \gamma$,  is formed in the near-edge regions
 with the widths  of the order of $ L_p $.
  The width $L_p$ strongly depend on the closeness
   to the viscous resonance [see Eq.~(\ref{pl_length_of_decay_exact})].
 The  viscoelastic perturbation is formed
  in the very vicinities of the edges with  the widths
of the order of  $ L_s = 1/\mathrm{Re}\lambda_2 $, $L_s \ll  L_p  $
[see Fig.~1(b) in the main text].
 The contribution of the viscoelastic contribution to the current $\mathbf{I}$
 is negligibility small [see Eq.~(\ref{V_x})].
 In the central region, $W/2-|y| \gtrsim \mathrm{Re}\lambda_1$,
 the linear response is  the Ohmic flow: it is formed
 just by the cyclotron motion of
 individual electrons and scattering on disorder.
  If the absorbtion power $ \mathcal{W} $
  is mainly determined by the central region, the response
  exhibits the usual cyclotron resonance
with a symmetric Lorentzian form exactly
at $\omega = \omega_c$ [see Fig.~4(a)].

At the frequencies  below the cyclotron resonance, $  \omega_c - \omega \gg \gamma$,
 the plasmonic-like nonoscillating  perturbation
  of the electron density and the hydrodynamic velocity
 are formed in the near-edge layers
 with the widths of the order of $ l_p \sim s/ \omega \ll  L_p  $.
Near the cyclotron resonance, $|\omega-\omega_c| \lesssim \gamma$,
 the plasmonic perturbations have the character lengths,
  $\widetilde{L}_p \sim s/\sqrt{\gamma \omega}$.
 As  we have assumed that $\gamma \to 0$, this value can be
 greater than the sample width $W$.

 In moderately  wide samples:
 \begin{equation}
 \label{wide}
  l_p  \ll W \ll L_p
  \:,
  \end{equation}
the plasmonic perturbations above the cyclotron resonance, $  \omega - \omega _c \gg \gamma$,
are formed inside the whole sample.   When the sample width
 is equal to an integer or a half-integer
number of the wavelengths of magnetoplamons,
   \begin{equation}
   \label{plasm_res}
   W= \frac{ m \pi }{q_p(\omega,\omega_c) }
   \:,
   \end{equation}
 the dependence $\mathcal{W}(\omega_c)$ exhibits
 the plasmonic resonances [see Fig.~4(b)].
 In Eq.~(\ref{plasm_res})  $m $ is an integer number
 and  $q_p(\omega,\omega_c)$ is the root of Eq.~(\ref{omega_p}).
 For a fixed $\omega$ the positions of the plasmonic resonance  frequency
 $\omega_c^{p,m} (W)$  can fall in the vicinity
 of the viscous resonance frequency $ \omega_c = \omega/2 $.
  In this case, the half-width of the plasmonic resonance,
  which is proportional to  the plasmon damping due to viscosity
  (\ref{Ips_gen}),   resonantly depends
  on the difference   $ \omega_c - \omega  /2 $
  [see Fig.~2(a) in the main text].

 In moderately wide samples [Eq.~(\ref{wide})]  the plasmonic part
 of the linear response below ($|\omega-\omega_c| \gg \gamma$)
 and near ($ \omega_c -\omega \lesssim\gamma$)
 the cyclotron resonance has the same properties
 as in the very wide samples [Eq.~(\ref{very_wide})].

{ \em b) Narrow samples: the behavior of the viscoelastic contribution. }
In the samples of medium widths:
 \begin{equation}
 \label{medium}
L_s  \ll W  \ll l_p
\:,
 \end{equation}
 where $L_s = 1/ \mathrm{Re} \, \lambda_2 $,  the viscoelastic part
 of the linear response, which is formed   in the near-edge regions
  with the widths   of the order of $L_s$, give a substantial contribution
 to the current $\mathbf{I}(t)$.
 In the central part of the sample, $W/2 - |y| \gg L_s $,
  the flow  is still controlled by magnetoplasmons.
The viscoelastic and the plasmonic contributions
to the current can be comparable.

 Above the viscous resonance, $\omega - 2\omega_c \gg 1/\tau_{ee}$
 the flow in the near-edge regions, $ W/2 - |y| \lesssim L_s $,
   have the form of standing waves of
     the transverse zero sound.  Below the viscous resonance,
      $ 2\omega_c - \omega  \gg  1/\tau_{ee}$,
  the viscoelastic  part of the linear response
  is located in the narrower near-edge layers
  with the widths $\sim l_s \ll L_s$
  and has the exponential non-oscillating  profile.
Due to the change of the character of the viscoelastic perturbation
 with the transition between these two regimes,
  the viscous resonance arises as a peak
   in the dependencies $\mathcal{W}(\omega_c)$
   and $I_{x,y} (\omega_c)$ [see Fig.~4(c)].
  From Eqs.~(\ref{I_x_gen}) and (\ref{lambda_2_in_visc_res})
  we immediately obtain for $\mathcal{W}(\omega_c) $
  near the viscous resonance:
 \begin{equation}
 \label{Visc_res_profile}
 \mathcal{W}(\omega_c)  \propto
 \mathrm{Re}\left(
  \frac{1}{ \sqrt{4 \omega_c^2 -\omega^2 -4i\omega /\tau _{ee}}}
   \right)
 \:.
 \end{equation}
 The peak  (\ref{Visc_res_profile}) is strongly asymmetric
 relative to the point $\omega_c=\omega/2 $.

Near the viscous resonance,
  $ |  \omega  - 2\omega_c | \lesssim 1/\tau_{ee}$,
the  viscoelastic perturbation has the form of
 the oscillations near the sample edges
 rapidly decaying in the direction inside the bulk
 with the same period and the decay length
 given by Eq.~(\ref{lambda_2_in_visc_res}).

In narrow samples,
 \begin{equation}
 \label{narrow}
  l_s \ll W \ll L_s
  \:,
  \end{equation}
above the viscous resonance,   $\omega - 2\omega_c \gg 1/\tau_{ee}$
the flow is controlled by viscosity in the whole sample.
 The distribution of the hydrodynamic velocity
  have the form of standing waves  of the transverse zero sound.
 When the sample width becomes equal to
  an integer or a half-integer number of wavelengths
   of the  transverse sound,
   \begin{equation}
   \label{zero_sound_res}
   W=  \frac{ m \pi }{ q_s(\omega,\omega_c) }
   \:,
   \end{equation}
   the resonances related to forming of the standing waves of the
    transverse sound occurs above the viscous resonance
     and exhibit themselves in
     $ \mathcal{W} (\omega_c)$ [see Figs.~4(c,d)].
  In Eq.~(\ref{zero_sound_res}) $m$ is an integer number and
 $  q_s(\omega,\omega_c)$ is the root of Eq.~(\ref{omega_v}).
  Such resonances are similar to the plasmonic resonances
  and the resonanses of any standing waves in a resonator.
  They can be named  ``the transverse zero
  sound resonances'' or ``magnetosound resonances''.

 In narrow samples [Eq.~(\ref{narrow})]   below the viscous resonance,
  $2\omega_c  - \omega  \gg 1/\tau_{ee}$,
  the viscoelastic part of the flow is
  non-oscillating exponential growth in the directions from the edges into the bulk and is
  located in the near-edge regions with the widths of the order od $l_s  $
  (similarly as in medium samples).

\begin{figure}[t!]
\centerline{\includegraphics[width=1.0\linewidth]{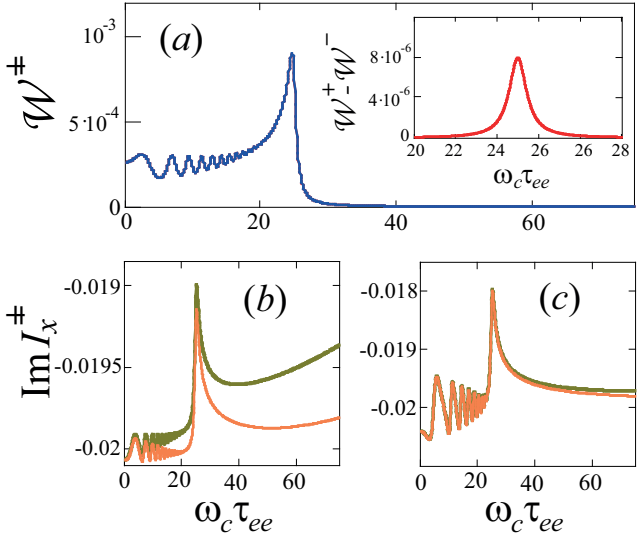}}
\caption{
Panel (a) shows the absorbtion  power
  $ W ^{\pm} (\omega_c) \approx E_0 \mathrm{Re} \, I ^{\pm} _x(\omega_c) $
 in units of $ \mathcal{\mathcal{W}}_0$
 for the case of the medium sample width ($W/l_{ee} = 1.8$)
 for the right (+) and left (-) polarizations of the ac fields.
  Inset on panel (a)
 shows the difference $\mathcal{W}^+(\omega_c) - \mathcal{W}^- (\omega_c)$
  near the plasmonic resonance, $\omega_c=\omega/2$.
  It is seen that the dependencies for the both  polarizations almost coincide.
 The parameters of the system are the same as in Fig.~4:
$\omega \tau_{ee} = 50$, $s/v_F = 300$, and
 $\gamma \tau_{ee} = 10^{-7}$.
 Panels (b) and (c) demonstrates the imaginary part of the ac current,
  $ \mathrm{Im} \, I^{\pm}_x(\omega_c) $,
 in units of $I_{x0} $
 for the medium sample widths [$W/l_{ee} = 1.8$ for panel (b)
 and  $W/l_{ee} = 0.8$ for panel (c)].
Orange and green curves correspond to  the right  and left  polarizations.
 }
\end{figure}

In maximally narrow samples,
 \begin{equation}
 W \ll l_s
 \:,
  \end{equation}
the viscoelastic contribution also dominates.  The flow
has a parabolic profile  in the whole sample
[see Eq.~(\ref{V_x}) in the limit $\lambda _{1,2} \to 0$].
This regime can be regarded as the ac Poiseuille flow
in magnetic field.  For the current we obtain from (\ref{I_x_gen}):
 \begin{equation}
  \label{I_narrow}
 I_x = \frac{e^2 n_0 E_0W^3 }{24\,  m \,\eta_0}
  \frac{  \omega^2 + 4 \omega_c^2
   + i \omega \tau_{ee}  ( - \omega^2 + 4 \omega_c^2 )  }
  { \omega^2  }
 \end{equation}
and $| I_y |  \ll  | I_x|$.
 We see from Eq.~(\ref{I_narrow}) that the viscous
 resonance  manifests itself  not in the current,
 but in the ac impedance $Z = E/I_x$.
 Near the resonance we have from Eq.~(\ref{I_narrow}):
 \begin{equation}
 \label{res_in_resistance}
  \mathrm{Re} \, Z \propto \frac{1}
  { 1 + ( \omega - 2 \omega_c ) ^2 \tau_{ee} ^2 }
  \:.
 \end{equation}

In the limit $\omega \tau_{ee} \gg 1$ the imaginary part
of the current (\ref{I_narrow}) is independent
on any relaxation parameter. The motion of the fluid
 in this regime resembles oscillations
 of a charged elastic media, glued near
 at the edges $y=\pm W/2$, in response of ac external field.
Herewith the real part of  Eq.~(\ref{I_narrow}) describes
 the relaxation of such elastic oscillations.
 By this way, for the maximally narrow samples,
 $ v_{F0} /\omega \ll W \ll l_{\omega}$, the viscoelasticity transition
 occurs at $\omega \tau_{ee} \sim 1$.

 By this way,  the viscous resonance manifests itself
  in different ways depending on the sample width $W$.
In wide samples it can be observed  by the resonant dependence
 of the half-width of the plasmonic resonance on $W$.
  In medium samples,  if the main contribution to the linear response
  comes from the viscoelastic part of the flow,
 the viscous resonance occurs due to changing of the character
  of the viscoelastic part of the flow, which is
  standing magnetosonic waves  at the ac frequencies above the viscous resonance
   and a non-oscillating decay below the viscous resonance.
   In this case, the    dependence $\mathcal{W}(\omega_c)$
   has the form of  an asymmetrical  peak near $\omega=2\omega_c$.
 In narrow samples when the standing waves of zero sound
 are well-formed, the magnetosound resonances related to
 coincidence of the sample width with an integer
 or half-integer number of wavelengths  of the transverse sound,
 appear in the dependence $\mathcal{W}(\omega_c)$.
In narrowest samples the ac Poiseuille flow
is formed and the viscous resonance manifest itself in the sample impedance.

An important feature of the viscoelastic part
of the linear response is its independence on the sign
 of the circular polarization of ac field [see Eq.~(\ref{V_x}].
With decrease the sample width,
 the dependence of $\mathbf{I}$ and $\mathcal{W}$
 on the sigh of the circular polarization becomes weaker
  and, finally, almost vanish for
  for both real and imaginary parts
  at the sample widths $W$ùà of the order of the decay length  of magnetosound,
  $L_s\sim l_{ee}$ (see Fig.~5).
 Herewith this dependence disappears faster  for
 the real part of the current $I_x$,
 than for its imaginary part than.

\section{ 3. Experimental results on non-linear ac  effects
 in high-mobility $\;\mathrm{GaAs}\;$ quantum wells
and  possible hydrodynamic  description of these effects }
  It seems very possible that  the viscous resonance is  responsible
for the giant peak in photoresistance  \cite{exp_GaAs_ac_1_,exp_GaAs_ac_2_}
 and  the peculiarities the photovoltaic effect \cite{exp_GaAs_ac_3_},
 that were  observed at $\omega = 2 \omega_c$
  in the high-mobility GaAs quantum wells.

  Indeed, first, it was stressed  in Ref.~\cite{exp_GaAs_ac_1_}
  that the peak in photoresistance
  and a very well pronounced giant negative magnetoresistance,
 explained in Ref.~\cite{je_visc_}
 as a manifestation of forming of a viscous flow,
  are observed in the {\em same best-quality} GaAs structures.
   If 2D electrons   in such structures form a viscous fluid,
    then any response of the system on ac field
  (absorbtion, photovoltage, photoresistance)
  must inevitably have  peculiarities
    at the frequency of the viscous resonance.

  Second, in the current paper we demonstrated
  that the peak of the viscous resonance
 can  be asymmetric as well as symmetric,
 depending on the sample width.
   Both these two forms of the peak are observed in
   the experiments \cite{exp_GaAs_ac_1_,exp_GaAs_ac_2_}
   in different samples at different conditions.

   Third, in Ref.~\cite{exp_GaAs_ac_3_} numerous
   magnetoplasmon resonances were observed
   in the photovoltage signal on  several GaAs
   high-quality structures of different geometries.
   The peculiarity at $\omega= 2 \omega _ c $ on the background of the ``forest''
   of plasmonic resonances was seen in the meander-gated samples.
   In such structures the
   space  inhomohenity of the flow and the relative role of viscosity
   are expected to be greater,
   compared with the other studied samples
   with flat gates and without any gates, where
    a smeared  peculiarity and   no peculiarity was observed at $\omega= 2 \omega _ c $ .

 Beside this, the following circumstances should be noted.
 It was experimentally shown in Refs.~\cite{Smet_1_,Smet_2_}
 that photoresistance and, in particular,
 the micro-wave induced resistance oscillations
  (MIRO) in high-mobility GaAs  quantum wells do not depend
  on the sign of circular polarization of radiation
 at $\omega_c \lesssim \omega$.
  At the same time, moderate negative magnetoresistance at low temperatures
  was observed \cite{Smet_2_}  on the sample studied
  in Ref.~\cite{Smet_1_,Smet_2_}
  (at higher temperatures a  positive magnetoresistance was observed).
  With increase of the frequency, the profile of the MIRO oscillations
   in the region $\omega_c  \approx \omega/2 $
  becomes more irregular (not sinusoidal) and exhibits
  some peak near $\omega_c \approx \omega/2$,
  as it was also observed  in a more pronounced
   manner in Ref.~\cite{exp_GaAs_ac_2_}.

   By this way, in Refs.~\cite{Smet_1_,Smet_2_}  a correlation
   between the appearance of a peak-like feature
    in photoconductivity near $\omega_c \approx \omega/2$
   and the independence of photoresistance
   on the sign of the circular polarization   was, possibly, established.
 In the current paper we demonstrate that the viscoelastic part
 of the linear response is independent
 on the sign of the circular polarization.
 These facts  could be an additional evidence
  that the peak at $\omega = 2 \omega_c $ in photoresistance
   of the high-mobility GaAs quantum wells
    is related to the viscous resonance.

 Oscillations of photoresistance of high-quality GaAs quantum wells
 were recently observed for the case of radiation
  in the terahertz diapason \cite{Ganichev_1_,Ganichev_2_}.
  The properties of the revealed oscillations
  are very similar to those of MIRO, in particular,
  they weakly depend or almost do not depend
 on the sign of the circular polarization of radiation.

 The assumption about a strong interaction between quasiparticles
   in  the 2D electron Fermi liquid in the high-mobility GaAs quantum wells
 seems to be fulfilled  because of the two following facts.
 First, the parameter $r_s$ that  characterize
  the strengths of the Coulomb interaction in  a 2D electron  system
  is about unity for the structures studied in
  Refs.~\cite{exp_GaAs_ac_1_,exp_GaAs_ac_2_,exp_GaAs_ac_3_}.
     Second, there are the experimental evidences
    that the effective mass of 2D electrons
    in high-mobility GaAs quantum wells
   is strongly renormalized due to
   inter-particle interaction \cite{book_,mass_}.
 A preliminary analysis \cite{future_}  demonstrates that
  a strong interaction between quasiparticles
  of the 2D electron  Fermi liquid justify
 the applicability of hydrodynamics   for the proper  description
 of  such  highly viscous fluids at the frequencies $\omega \gg 1/\tau_{ee}$.

 To construct the theories of the photoresistance
  and the photovoltaic effects, one should take into account
  some nonlinear terms   in  the hydrodynamic equation
  (\ref{Navier_Stocks_B_with_time_disp}) following, for example,
  Refs.~\cite{Lifshits_Dyakonov_,Beltukov_Dyakonov_}.
  In these works  the theories   of the photovoltaic effect
  and the  MIRO oscillations in photoresistance
  were developed for 2D electrons in a disordered  systems.
  One of the main elements of
  the theories~\cite{Lifshits_Dyakonov_,Beltukov_Dyakonov_}
  is the linear response of free 2D electrons on ac field:
$ \mathbf{V}_0(y)=eE_0 \mathbf{A}_0/ (2m) $.
   In the theory   of the photoresistance and the photovoltage effects
  in  a  viscous electron fluid  in a sample  without disorder,
  the linear response~(\ref{V_x})
  should be used instead  of $\mathbf{V}_0(y)$.

All the previous theories of the MIRO effect  were developed
for the model of 2D electrons in a disordered system \cite{rev_M_}.
 These theories explained very well the shape
 of photoresistance oscillations and their dependence
 on temperature and other parameters.
The main problem of all the theories was the inconsistency
between the lack of dependence of the effect
 on the sign of the circular polarization
 in experiment and the presence of such a dependence
  in  theories. As we now  have the evidences that
  2D electrons in high-mobility GaAs structures
 form a highly viscous fluid (``electron honey''),
 it is logical to construct the theory of the MIRO effect
   for the case of such fluid flowing    in a sample
   without any disorder.

   In real samples, apparently, there exist
    the two contributions to photoresistance and other kinetic effects:
   the part of the signal is controlled by scattering
   of individual electrons   on disorder and the other part
   is   controlled by forming the electron fluid, being
   is independent on the sign of the circular polarization.

   The theory~\cite{Beltukov_Dyakonov_} takes into account
   the memory effects  in scattering of electrons
   on smooth localized defects  in magnetic field.
   This lead to a proper description of     the MIRO oscillations
   except the explanation of
   their independence the sign of the circular polarization.
 Possibly, taking into account the memory effects
 in {\em electron-electron scattering}  in the hydrodynamic theory
  of nonlinear magnetotransport will lead not only to explanation
  of the giant peak in photoresistance,   but also
  to a proper description of the MIRO oscillations, similar
  to the description within the models for 2D independent electrons in samples with bulk disorder.
  The memory effects in interparticle collisions
  were considered for the case of Boltzmann gas
  in Refs.~\cite{mem_old_1_,mem_old_2_}
  and, very recently, for hydrodynamic transport of electrons
  in graphene in Ref.~\cite{mem_recent_}.
   In the future possible theory of hydrodynamic magnetotransport
  taking into account the memory effects
   in electron-electron collisions in magnetic field,
    the calculated linear response
  (\ref{V_x}) will be an element, which will be used instead of the linear
  response of free electrons.
 Thus  the resulting photoresistance at magnetic fields
 $\omega_c < \omega$
  will not depend on the sign
 of the circular polarization for the narrow samples.


\begin{thebibliography}{}
\bibitem{comm}
 For the case of zero magnetic field, the expression similar
to Eqs.~(\ref{V_x}) and (\ref{I_x_gen}) were independently derived in Ref.~\cite{recentest,recentest2}.
\end{thebibliography}
\end{document}